# Clinical Reasoning AI for Oncology Treatment Planning: A Multi-Specialty Case-Based Evaluation

Philippe E. Spiess*[,1], MD, MS, FACS, FRCS(C); Md Muntasir Zitu*[,1], PhD; Alison Walker[1], MD; Daniel A. Anaya[1], MD; Robert M. Wenham[1], MD; Michael Vogelbaum[1], MD, PhD; Daniel Grass[1], MD; Ali-Musa Jaffer[1], MD; Amod Sarnaik[1], MD; Caitlin McMullen[1], MD; Christine Sam[1], MD; John V. Kiluk[1], MD; Tianshi Liu[1], MD; Tiago Biachi[1], MD; Julio Powsang[1], MD; Jing-Yi Chern[1], MD; Roger Li[1], MD; Seth Felder[1], MD; Samuel Reynolds[1], MD; Michael Shafique[1], MD; Alison Sheehan[1], APP; Ashley Layman[1], APP; Cydney A. Warfield[1], APP; Derrick Legoas[1], APP; Jaclyn Parrinello[1], APP; Jena Schmitz[1], APP; Kevin Eaton[1], APP; Mark Honor[1], APP; Luis Felipe[1], PhD; Issam ElNaqa[1], PhD, DABR; Elier Delgado[2,3], MSc; Talia Berler[1], MSc; Rachael V. Phillips[3], PhD; Frantz Francisque[4], MD; Carlos Garcia Fernandez[1,3], PhD; Gilmer Valdes[1,3], DABR

[1] Moffitt Cancer Center and Research Institute, 12902 Magnolia Drive, Tampa, FL 33612, USA.

[2] Innova Montréal Inc., IT Services and Consulting, Montréal, QC, Canada.

[3] Oncobrain, Inc., Tampa, FL, USA.

[4] Advanced Cancer Treatment Centers, 15211 Cortez Boulevard, Brooksville, Florida 34613, USA.

*Philippe E Spiess, MD, and Md Muntasir Zitu, PhD, contributed equally as co–first authors.

Corresponding author: Md Muntasir Zitu, PhD
Moffitt Cancer Center, 12902 Magnolia Drive, Tampa, FL 33612, USA
Tel: +1-317-985-4809, Email: mdmuntasir.zitu@moffitt.org

## Abstract

**Background:** A persistent survival disparity separates academic centers from the community settings where more than 80% of U.S. cancer care is delivered. This gap is exacerbated by an accelerating cognitive burden on clinicians, who must integrate complex genomics, evolving staging, radiologic and pathologic findings, and frequent guideline updates. Existing oncology decision-support approaches, including static guideline-reference tools, rule-based clinical decision support, and general-purpose LLM interfaces, often lack the nuance or safety required for high-stakes oncology planning. We sought to evaluate an artificial intelligence (AI) clinical reasoning platform, OncoBrain, for oncology treatment-plan generation as an early step toward Oncology General Intelligence (OGI).

**Methods:** OncoBrain is a decision-support platform that orchestrates general-purpose LLMs with (i) a cancer-specific graph retrieval–augmented generation (Graph RAG) layer, (ii) a gold-standard treatment-plan corpus as "long-term memory," and (iii) a model-agnostic safety layer (CHECK) for hallucination detection and suppression. We conducted a multi-component evaluation using clinician-enriched case summaries spanning gynecologic, genitourinary, neuro-oncology, gastrointestinal/hepatobiliary, and hematologic malignancy scenarios: cases were drafted with the most capable GPT-family model available at the time of generation and then edited by clinicians to better reflect real-world scenarios during an iterative clarification ("treatment-plan") cycle. Three clinician groups contributed structured evaluations of 173 cases in total using a common 16-item instrument spanning scientific accuracy, safety, workflow integration, adoption/overall value, and free-text feedback: subspecialist oncologists completed 50 vignette reviews, additional physician reviewers (MDs) completed 78 case-level reviews, and advanced practice providers (APPs) completed 45 case-level reviews. We also describe an illustrative de-identified real-patient case experience from a community setting.

**Results:** Across the three clinician groups (subspecialist oncologists, MDs, and APPs), ratings were highest for scientific accuracy, evidence support, and safety, with more variable scores for workflow integration and time savings. On a 5-point scale (1 = least favorable, 5 = most favorable), mean scores for alignment with evidence and guidelines were 4.60 (subspecialist oncologists), 4.56 (MD reviewers), and 4.70 (advanced practice providers); mean scores for no safety or misinformation concerns were 4.80, 4.40, and 4.60, respectively. Workflow integration averaged 4.50, 3.94, and 4.00, and perceived time savings averaged 5.00, 3.89, and 3.60.

**Conclusions:** In this multi-specialty, vignette-based evaluation, OncoBrain generated oncology treatment plans that subspecialist oncologists judged to be guideline-concordant, clinically acceptable, and easy to supervise within existing workflows. These findings suggest that a carefully engineered AI reasoning platform, combining domain-specific retrieval, expert-derived long-term memory, and a dedicated safety layer, can meaningfully support treatment-plan generation and offer a concrete step toward OGI. Prospective studies using real-world data and deployment in community settings will be essential to determine the impact on care patterns and equity.

## Introduction

A substantial gap in cancer survival persists between academic medical centers and community settings, where more than 80% of patients in the United States receive care. [1-4] Population-based analyses confirm that patients treated at high-volume or NCI-designated centers experience superior survival, an advantage driven in part by greater access to subspecialty expertise and multidisciplinary care [5-7]. This 'evidence-to-practice' gap is widening as the cognitive load on clinicians increases; oncologists must integrate genomics, sub-staging, molecular-clinical disease stratification, advanced radiologic findings, timing of surgery and systemic therapy, and clinical trial options from clinical guidelines and evidence sources that are revised multiple times per year [8-11]. Existing oncology decision-support tools, including static guideline-reference resources and conventional rule-based clinical decision support (CDS), fail to bridge this divide. Conventional rule-based clinical decision support (CDS) cannot manage this combinatorial complexity, while unconstrained general-purpose large language models (LLMs) are unreliable, prone to hallucination, and lack provenance, making their ad hoc use for high-stakes decisions unsafe [12-15].

Addressing these challenges requires not another static guideline app or unconstrained chatbot, but an AI clinical reasoning partner built from the ground up for oncology. Conceptually, we frame this goal as building Oncology General Intelligence (OGI): the long-term aim of developing systems that can support the full spectrum of oncologist cognitive tasks, such as staging and biomarker interpretation, treatment plan generation, toxicity-aware regimen selection, and clinical trial matching, one well-defined task at a time. We argue that OGI is likely to be achievable well before artificial general intelligence (AGI), because the set of clinically meaningful oncologist tasks is finite and relatively well defined, whereas AGI must span an essentially unbounded task space across domains. Within this finite space, treatment-plan generation is a natural first task. It is among the most difficult and consequential cognitive activities in oncology, demanding integration of heterogeneous data and nuanced trade-offs between efficacy, toxicity, logistics, and patient goals, as well as providing a structured, evaluable endpoint: concrete plans that can be compared against guidelines and expert opinion.

Our approach draws inspiration from a field that has grappled with high-stakes planning for decades: radiation oncology. In radiotherapy, auditable physician-in-the-loop systems with optimization engines generate candidate plans, quantify trade-offs, and document constraints and rationale before a clinician signs off [16-18]. We extend this philosophy, such as traceable reasoning, rigorous auditing, and stepwise clinical approval, from a single modality to oncology-wide treatment planning, aiming to optimize the therapeutic index across systemic, surgical, and radiation therapies. In this paradigm, AI is not an oracle but an assistant that proposes a small set of optimized, guideline-anchored treatment strategies along a "Pareto frontier," meaning a set of options reflecting different trade-offs between efficacy and toxicity, with the clinician selecting and refining the final plan.

To operationalize this vision, we developed an AI clinical reasoning platform (hereafter referred to as the OncoBrain) that combines three key components [19]. First, a cancer-specific graph retrieval–augmented generation (Graph RAG) module organizes knowledge as a domain-aware graph linking diseases, stages, biomarkers, drugs, regimens, toxicities, and clinical trials, overcoming the limitations of naive vector search over unstructured text. Second, a gold-standard treatment-plan corpus derived from a high-performing cancer center provides a "long-term memory" of expert decisions; this corpus is used both for in-context guidance ("what would an expert do for a case like this?") and for automated validation of candidate plans. Third, a model-agnostic safety layer (CHECK) conducts a final "physics check" on LLM-generated outputs, using distributional signals and independent classifiers to detect hallucinations and unsafe recommendations before any plan is shown to the clinician, building on prior work characterizing medical hallucination phenotypes [19].

In this study, we evaluate the OncoBrain AI clinical reasoning platform for oncology treatment plan generation. Using specialty-specific cancer cases reviewed by expert oncologists, we assess the platform's ability to propose guideline-concordant, clinically acceptable treatment plans; characterize its safety profile, including the presence or absence of clinically significant errors; and explore clinicians' perceptions of trust, usability, and workflow fit. By situating this evaluation within the broader OGI roadmap and the longstanding equity gap between community and academic cancer care, we aim to understand whether a rigorously engineered AI reasoning partner can serve as a credible first step toward democratizing access to expert-level oncology decision support. We also made the platform accessible via a registration-based access process [20].

## Methods

We conducted a multi-component evaluation comprising (1) a specialty-specific synthetic vignette assessment by subspecialist oncologists, (2) an expanded clinician feedback evaluation including physicians (MDs), (3) an advanced practice providers group (APPs), and (4) an illustrative real-patient case experience in a community setting.

### Platform overview

Our OncoBrain is an AI decision-support platform that orchestrates general-purpose large language models (LLMs) as a model-agnostic inference layer for multi-step oncology planning, rather than operating as a separately trained oncology-specific foundation model. The OncoBrain uses RAG to query curated sources such as ASCO and NCCN Guidelines®, Drugs@FDA, PubMed, ClinicalTrials.gov, and genomic databases (e.g., OncoKB, CIViC), to produce citable, evidence-based rationale. The platform does not bypass publisher or society access controls; rather, it retrieves from publicly available content and authorized sources. Additionally, it uses test-time compute scaling in agentic workflows to improve the precision and accuracy of the answers. OncoBrain's safety mechanisms include: (1) physician-in-the-loop checkpoints; (2)

automated classifiers flagging ambiguous, contraindicated, or harmful outputs; and (3) test-time compute increase for complex cases.

In the current implementation, OncoBrain exposes three linked transparency features that were available to clinician review during this evaluation: a Sources panel, a Clinical Reasoning view, and a Safety Score panel. (Supplementary section 3.0; Supplementary Figures 1-3). A Sources panel anchors each recommendation to specific guideline passages, FDA labels, and other authoritative references, enabling rapid spot verification of citations. A Clinical Reasoning view presents a stepwise trace from query to retrieved evidence to final recommendation, with each step linked to its underlying source, creating an auditable path from input to output. A Safety Score panel summarizes a multi-domain rubric (e.g., misinformation, unsafe medication guidance, potential patient harm, privacy/ethics) on a 1–5 scale (higher scores indicating greater safety) for each case. Together, these features are intended to support efficient validation and documentation while preserving clinicians-in-the-loop during plan development. The platform is accessible through a registration-based process.

**Case generation and clinician workflow**

At the initial stage, synthetic case summaries were generated using the most capable OpenAI GPT model available at the time of case creation (GPT-4 for the subspecialty vignette set; GPT-5-series models for the structured clinician feedback cohorts), using the same prespecified prompting framework (Supplementary section 1.0). ChatGPT generated vignettes mirroring tumor board presentations (initial presentation, diagnostic work-up, tissue diagnosis, staging, and molecular/mutational analysis). Next, the expert evaluators were instructed to edit those cases as needed to ensure that the vignettes reflected plausible real-world clinical scenarios (Figure 1). Finally, OncoBrain was provided with those cases as input, and it iteratively posed clarifying questions to resolve missing clinical details. Once the case content was finalized, the platform generated a treatment plan using retrieval-augmented generation with citable sources (e.g., guidelines, Drugs@FDA, PubMed abstracts, and CIViC). For the illustrative real-patient case experience, a clinician prepared a de-identified case summary from routine community practice and followed the same clarification and plan-generation workflow.

**Evaluation and domains**

Across all clinician evaluator groups, we used a common structured feedback framework to assess OncoBrain's oncology treatment plan generation. Clinicians received brief training, used the platform's clarification workflow as needed, and then reviewed an OncoBrain-generated plan that included inline citations and an explicit reasoning trace. Each plan was evaluated using the same 16-item instrument spanning five domains (Table 1): (1) Scientific Accuracy & Clinical Utility; (2) Safety & Medicolegal Confidence; (3) Workflow Integration & Efficiency; (4) Adoption/overall Value; and (5) Free-text query. Domain 1 assessed guideline/evidence concordance and clinical appropriateness; Domain 2 assessed safety signals, contraindications, and medicolegal comfort; Domain 3 assessed workflow fit, time saved, and ease of review, correction, or override; Domain 4 assessed Adoption/overall value; and Domain 5 captured free-text feedback on unique value, limitations, and requested improvements. For each questionnaire item, we report the number of non-missing case-level responses. Because clinicians could omit items they considered not applicable, question-level summaries were calculated using available responses only. Item completion was high overall, with complete responses for the structured rating items in the subspecialist cohort and modest item-level missingness in the MD and APP groups, as detailed by question-level N values in Table 1.

Three clinician groups contributed structured evaluations. Five board-certified surgical/medical oncologists (gynecologic, genitourinary, neurosurgical-oncology, gastrointestinal/hepatobiliary, and malignant hematology), each with >10 years' experience, evaluated clinician-enriched synthetic vignettes within their subspecialty domain, yielding 50 case-level reviews (one subspecialist review per vignette). To broaden evaluator representation, we additionally collected structured feedback from physicians (MDs; *n*=14) and advanced practice providers (APPs; *n*=8), who reviewed variable numbers of cases in a self-paced manner and contributed 78 and 45 case-level reviews, respectively. In total, clinicians completed 173 case-level evaluations across cohorts (50 subspecialist vignette reviews, 78 MD reviews, and 45 APP reviews). We also describe an illustrative de-identified real-patient case experience from a community setting. Please see Supplementary section 2.0 for a sample OncoBrain-clinician transcript.

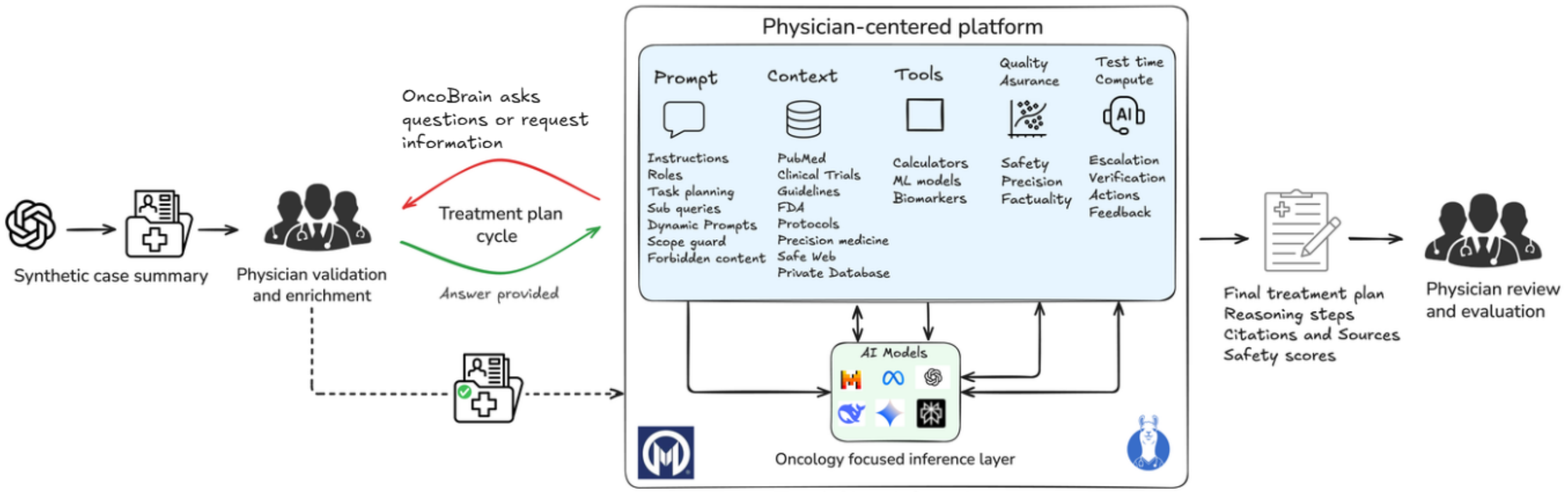

Figure 1: OncoBrain evaluation workflow for treatment plan generation. Synthetic case summaries are first reviewed and enriched by clinicians, who can iteratively answer OncoBrain clarifying questions in a "treatment-plan cycle." The validated case then enters a physician-centered workspace that organizes the interaction into five elements: (1) a **Prompt** layer that encodes roles, task instructions, and guardrails; (2) a **Context** layer that retrieves oncology knowledge from curated sources such as guidelines, clinical trials, PubMed, FDA documents, institutional protocols, and private databases (distinct from the gold-standard treatment-plan corpus, which reflects expert decisions rather than primary sources); (3) a set of **Tools** (for example, calculators, models, and biomarker utilities) that can be invoked as needed; (4) a **Quality-assurance** layer focused on safety, precision, and factuality; and (5) **Test-time compute**, which governs escalation, verification actions, and feedback.

## Results

### Quantitative toplines

Across clinician groups, ratings were highest for scientific accuracy, evidence support, and safety, with more variable scores for workflow integration and time savings. On a 5-point scale (1-5, higher scores indicate more favorable ratings), mean scores for matched evidence and guidelines were 4.60 in subspecialist oncologists (50 vignette reviews), 4.56 in MD reviewers (78 case-level reviews), and 4.70 in advanced practice provider (APP) reviewers (45 case-level reviews). Mean ratings for no safety/misinformation concerns were 4.80, 4.40, and 4.60, respectively; smooth workflow integration averaged 4.50, 3.94, and 4.00; and perceived time savings averaged 5.00, 3.89, and 3.60. The number of feedback for question-level responses varied because clinicians could omit questions they found non-applicable, and because not all assigned case reviews were completed. Table 1 summarizes item-wise distributions and qualitative themes across cohorts.

**Table 1**. Clinician evaluation of OncoBrain treatment plans across clinician groups. Items Q1-Q12 were rated on a 5-point Likert scale (1 = least favorable, 5 = most favorable). For each question, values are calculated only with the number of valid case-level responses (i.e., the value of N). Results are grouped as: Section 1, scientific accuracy & guideline alignment (Q1-Q4); Section 2, safety & medicolegal confidence (Q5-Q7); Section 3, workflow integration & efficiency (Q8-Q10); Section 4, adoption/overall value (Q11-Q12). Section 5 summarizes qualitative themes from free-text responses (Q13-Q16). *med stands for median*

| **Section & Item** | **Subspecialist oncologists (50 case reviews)** | | | | | **Physicians (MD) (78 case reviews)** | | | | | **Advanced practice providers (45 case reviews)** | | | | |
|---|---|---|---|---|---|---|---|---|---|---|---|---|---|---|---|
| | min | max | mean | med | **N** | min | max | mean | med | **N** | min | max | mean | med | **N** |
| Section 1 - Scientific Accuracy & Clinical Utility | | | | | | | | | | | | | | | |
| Q1. Matched evidence & guidelines | 4 | 5 | 4.60 | 5 | 50 | 2 | 5 | 4.56 | 5.0 | 78 | 2 | 5 | 4.7 | 5 | 45 |
| Q2. Sound & patient-appropriate explanations | 4 | 5 | 4.60 | 5 | 50 | 2 | 5 | 4.69 | 5.0 | 78 | 2 | 5 | 4.7 | 5 | 45 |
| Q3. Deferred to clinician judgment | 4 | 5 | 4.60 | 5 | 50 | 2 | 5 | 4.6 | 5.0 | 77 | 2 | 5 | 4.7 | 5 | 45 |
| Q4. Reliable citations & evidence | 4 | 5 | 4.75 | 5 | 50 | 1 | 5 | 4.28 | 5.0 | 75 | 2 | 5 | 4.7 | 5 | 45 |
| Section 2 - Safety & Medicolegal Confidence | | | | | | | | | | | | | | | |
| Q5. No safety/misinformation concerns | 4 | 5 | 4.8 | 5 | 50 | 1 | 5 | 4.4 | 5.0 | 78 | 2 | 5 | 4.6 | 5 | 45 |
| Q6. I was comfortable with the AI tool's guardrails for complex or ambiguous situations. | 3 | 5 | 4.4 | 5 | 50 | 2 | 5 | 4.41 | 5.0 | 78 | 2 | 5 | 4.6 | 5 | 45 |
| Q7. No medicolegal concerns | 4 | 5 | 4.75 | 5 | 50 | 1 | 5 | 4.45 | 5.0 | 78 | 2 | 5 | 4.6 | 5 | 45 |
| Section 3 - Workflow Integration & Efficiency | | | | | | | | | | | | | | | |
| Q8. Smooth workflow integration | 3 | 5 | 4.5 | 5 | 50 | 1 | 5 | 3.94 | 4.0 | 77 | 1 | 5 | 4 | 4 | 38 |
| Q9. Saved time | 5 | 5 | 5 | 5 | 50 | 1 | 5 | 3.89 | 4.5 | 76 | 1 | 5 | 3.6 | 4 | 40 |
| Q10. I found it easy to review, correct, or override AI content before sharing with patients. | 5 | 5 | 5 | 5 | 50 | 1 | 5 | 4.21 | 5.0 | 75 | 1 | 5 | 4.2 | 5 | 41 |
| Section 4 - Adoption/overall value | | | | | | | | | | | | | | | |
| Q11. I would recommend this AI tool to colleagues. | 5 | 5 | 5 | 5 | 50 | 2 | 5 | 4.22 | 5.0 | 69 | 2 | 5 | 4.1 | 5 | 41 |
| Q12. Added meaningful value | 5 | 5 | 5 | 5 | 50 | 1 | 5 | 4.26 | 5.0 | 69 | 2 | 5 | 4.5 | 5 | 41 |
| Section 5 - Free-text query | Response | | | | | | | | | | | | | | |
| Q13. Unique values of OncoBrain vs general-purpose GenAI tools | Clinicians emphasized oncology-focused synthesis, guideline-linked citations, and surfacing relevant clinical trials; several compared outputs with OpenEvidence/Perplexity but described OncoBrain as more plan-oriented. | | | | | | | | | | | | | | |
| Q14. Practice benefits | Benefits included rapid access to guideline-aligned options and trial considerations, concise literature overviews for complex decisions, and support for patient communication and documentation. | | | | | | | | | | | | | | |
| Q15. Limitations/risks | Limitations included response latency or verbosity, dependence on input completeness, occasional disagreements with clinician preferences, and occasional citation mismatches when using follow-up questions outside standard plan generation. | | | | | | | | | | | | | | |

| | |
|---|---|
| Q16. Requested improvements | Requested improvements included faster outputs, streamlined interface and workflow integration (including EHR integration), more proactive prompts for key biomarkers and operability/anatomic criteria, and expanded capabilities (e.g., imaging interpretation and error checking). |

**Thematic free-text synthesis**

Across cohorts, free-text responses converged on four themes. a) Unique value vs general-purpose GenAI tools: clinicians emphasized oncology-focused synthesis, guideline-linked citations, and the ability to surface relevant clinical trials; several compared outputs with tools such as OpenEvidence or Perplexity and described OncoBrain as more plan-oriented. b) Practice benefits: clinicians cited rapid access to guideline-aligned options, concise literature overviews for complex decisions, and support for patient communication and documentation. c) Limitations/risks: clinicians noted response latency or verbosity, dependence on input completeness, and occasional disagreements with clinician preferences; a minority reported citation mismatches when using free-text follow-up questions outside standard treatment-plan generation. d) Requested improvements: clinicians requested faster outputs, streamlined interface and workflow integration (including EHR integration), more proactive prompts for key biomarkers and operability/anatomic criteria, and expanded capabilities (e.g., imaging interpretation and error checking). Figure 2 illustrates the word-cloud build from the free-text responses.

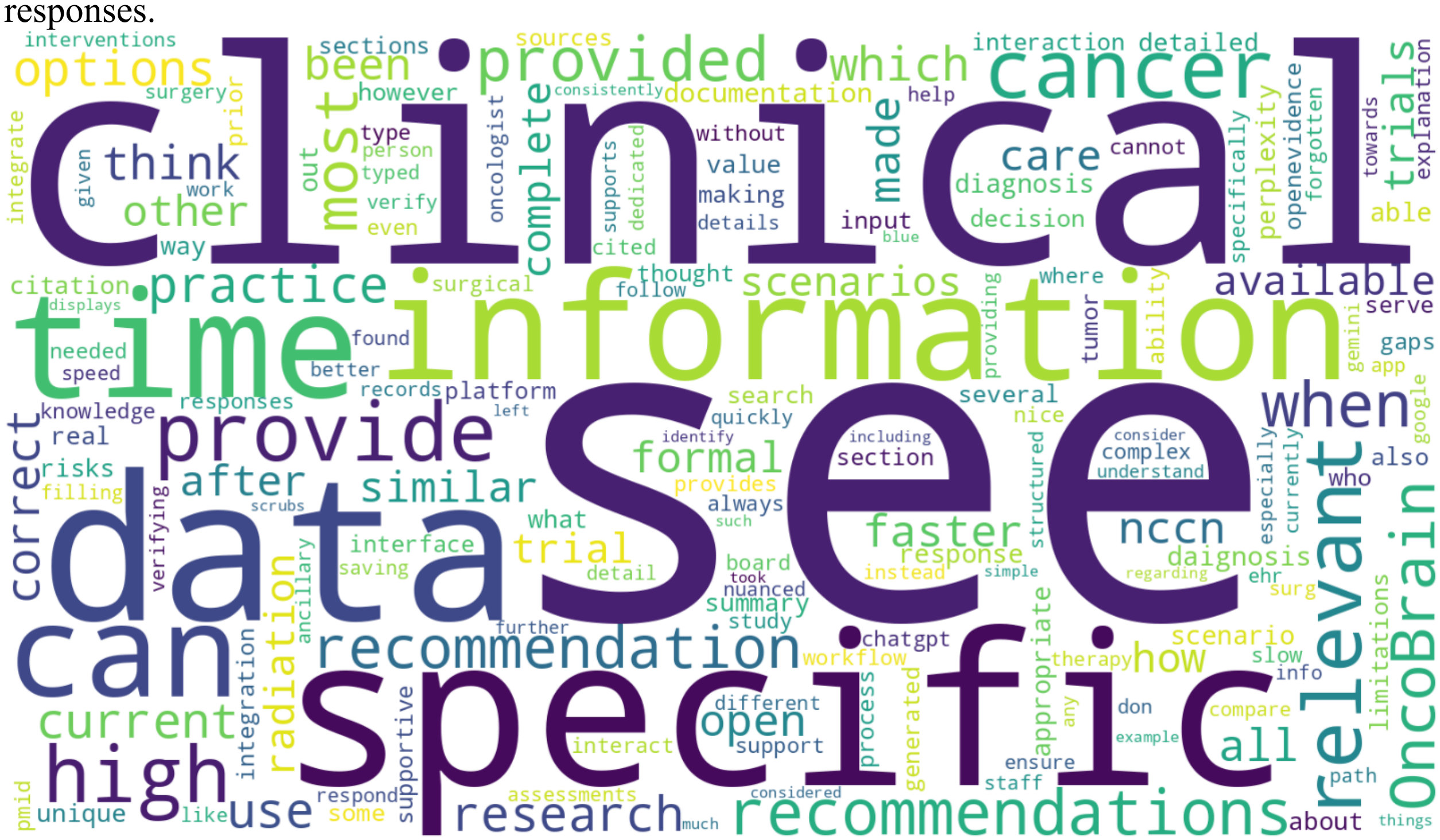

Figure 2. Word cloud of clinician free-text feedback (Q13–Q16). All open-ended responses for Q13–Q16 were pooled across evaluation forms. Text was lowercased, tokenized into unigrams, and common stopwords plus generic survey filler terms were excluded.

**Illustrative real-patient case experience**

We include a de-identified illustrative case from routine community oncology practice to show how the platform may support treatment selection in a complex real-world setting, in which a community clinician used OncoBrain to assist with management of a complex patient.

Case: An 80-year-old man was described with cirrhosis-associated hepatocellular carcinoma, previously treated with definitive radiation therapy, iron deficiency anemia, History of prostate cancer s/p SBRT who was followed for progressive cutaneous T-cell lymphoma. Despite phototherapy, his skin lesions progressed, and biopsy demonstrated epidermotropic T-cell lymphoma with an atypical immunophenotype (CD3-positive; CD4-, CD8-, CD5-, CD30-negative), raising concern for a more aggressive peripheral T-cell process. PET CT and peripheral flow cytometry showed no systemic disease. Treatment selection in the context of advanced age, cirrhosis, and prior HCC was a challenging process.

OncoBrain feedback summary: OncoBrain synthesized comparative data on efficacy, toxicity, treatment duration, response rates, and survival across available therapies, including methotrexate, bexarotene, interferon, histone deacetylase inhibitors, and mogamulizumab, highlighting higher response rates with bexarotene and interferon and more modest responses with HDAC inhibitors. After reviewing the data with the patient and family, the patient elected to proceed with bexarotene, which the patient tolerated overall well on early follow-up with hematologic stability and no hepatic toxicity. This case illustrates how oncology-specific, AI-validated decision-support systems can assist community clinicians in comparative treatment selection, balancing efficacy, toxicity, and patient-centered goals in complex clinical scenarios.

## Discussion

In this multi-specialty, multi-cohort evaluation of AI-generated oncology treatment plans, the OncoBrain AI clinical reasoning platform for oncology treatment-plan generation produced strategies that expert oncologists generally judged to be guideline-concordant, clinically acceptable, and safe across a diverse set of synthetic but realistic cancer vignettes. These vignettes were initially generated using ChatGPT and then reviewed by disease-specialized physicians to ensure clinical plausibility and alignment with real-world decision contexts. Clinicians reported that the generated plans were aligned with current standards, supported by authoritative citations, and straightforward to review, correct, or override. The consistently high median ratings and narrower spread among subspecialist oncologists likely reflect the study design, in which these reviewers evaluated clinician-enriched cases within their own disease domains, whereas the broader MD and APP groups contributed more heterogeneous, self-paced reviews and therefore showed greater variability, particularly for workflow integration and perceived time savings. These findings suggest that a carefully engineered AI reasoning partner can shoulder a meaningful portion of the cognitive work in treatment planning without displacing the oncologist's central role.

A distinctive aspect of this work is treating *treatment-plan generation itself* as the primary task, rather than as a by-product of short-form question answering. Here, the platform was evaluated on its ability to propose a complete, multi-step plan with explicit rationale, and to do so in a way that experts deem consistent with current practice. Within the OGI roadmap, these results support a “one task at a time” approach, starting with a high-value task central to clinical work. This evaluation of a full-stack AI clinical reasoning platform, not a standalone model, prototypes generalizable design patterns that may generalize beyond oncology: a cancer-specific graph-based RAG, a "long-term memory" built from expert decisions, and a model-agnostic safety layer that can sit atop different foundation models as they evolve.To support external use, the platform is publicly accessible through a registration-based access process.

Although our study does not address patient outcomes or system-level inequities directly, it has implications for the *equity questions*. The ability to generate expert-acceptable plans with transparent rationale is a necessary precondition for using such a system to support generalists in settings without ready access to subspecialty tumor boards. If future prospective work confirms similar performance in community environments, systems of this type could help narrow structural gaps in access to expert-level reasoning, even if they cannot solve broader resource disparities.

This study also highlights the importance of *human–AI interaction design* in high-stakes settings. Clinicians engaged in a “treatment-plan cycle” where the system asked clarifying questions and the physician filled gaps before the platform generated a plan. This process was perceived not as a burden but as an extension of the usual mental steps for preparing a detailed consultation. The finding that the platform’s outputs were easy to review, edit, and document, preserving medicolegal comfort and a clear locus of responsibility, reinforces that technical performance alone is insufficient; systems must be designed so that clinicians can safely and efficiently *work with* the AI.

Several limitations must be acknowledged. The evaluation used clinician-enriched synthetic vignettes rather than live electronic health records because this study was designed as a controlled evaluation of treatment-plan quality, and platform performance may vary with ambiguous or "messy" real-world data. The modest sample size was sufficient for these descriptive insights but not for formal hypothesis testing, and the work focuses on plan quality and user perceptions rather than downstream clinical outcomes. Prospective studies in community and academic settings will be required to determine whether AI-assisted treatment planning meaningfully changes care patterns or patient outcomes. Clinically, evaluations should expand into community oncology practices, and from a systems perspective, tighter integration with electronic health records and robust governance frameworks will be crucial for real-world adoption.

## Conclusion

In summary, this study provides a broad, multi-cohort clinician evaluation of AI-generated oncology treatment plans and suggests that an AI clinical reasoning platform, engineered with domain-specific retrieval, expert-derived long-term memory, and a dedicated safety layer, can generate plans that oncologists judge to be guideline-concordant, clinically acceptable, and usable within their workflows. Within the broader vision of OGI, these results suggest that starting with a single, high-value task and designing around safety, transparency, and physician oversight can yield tools that clinicians are willing to engage with in high-stakes settings. Realizing the full potential of such systems will require rigorous prospective evaluation and sustained collaboration, but the trajectory is clear: AI can move from generic "chatbots" toward accountable, clinician-guided partners in cancer care.

## Platform Availability

OncoBrain is accessible via registration at https://www.oncobrain.ai/. Access is governed by the platform's authorization and oversight processes.

## Acknowledgments

We thank Marcia Amnay for her valuable assistance in facilitating communication and coordination throughout the development of this study.

## Conflicts of Interest

The authors declare no conflicts of interest.

## Funding

This research received no external funding.